\documentclass[aps,pra,twocolumn,reprint,noeprint,superscriptaddress,amsmath,amssymb]{revtex4-1}
\usepackage{graphicx}
\usepackage{bm}
\usepackage{color}
\usepackage{subfigure}
\usepackage{bbm}
\usepackage{amsmath}
\usepackage[colorlinks, linkcolor=blue, urlcolor=blue, citecolor=blue]{hyperref}

\begin{document}
\title{Nonlinear interferometry beyond classical limit facilitated by cyclic dynamics}
\author{Qi Liu}
\thanks{These authors contributed equally to this work.}
\affiliation{State Key Laboratory of Low Dimensional Quantum Physics, Department of Physics, Tsinghua University, Beijing 100084, China}
\author{Ling-Na Wu}
\thanks{These authors contributed equally to this work.}
\affiliation{Institut f\"ur Theoretische Physik, Technische Universit\"at Berlin,\\
Hardenbergstra\ss e 36, Berlin 10623, Germany}
\author{Jia-Hao Cao}
\affiliation{State Key Laboratory of Low Dimensional Quantum Physics, Department of Physics, Tsinghua University, Beijing 100084, China}
\author{Tian-Wei Mao}
\affiliation{State Key Laboratory of Low Dimensional Quantum Physics, Department of Physics, Tsinghua University, Beijing 100084, China}
 \author{Xin-Wei Li}
\affiliation{State Key Laboratory of Low Dimensional Quantum Physics, Department of Physics, Tsinghua University, Beijing 100084, China}
\author{Shuai-Feng Guo}
\affiliation{State Key Laboratory of Low Dimensional Quantum Physics, Department of Physics, Tsinghua University, Beijing 100084, China}
\author{Meng Khoon Tey}
 \affiliation{State Key Laboratory of Low Dimensional Quantum Physics, Department of Physics, Tsinghua University, Beijing 100084, China}
 \affiliation{Frontier Science Center for Quantum Information, Beijing 100193, China}
\author{Li You}
\email[]{lyou@mail.tsinghua.edu.cn}
 \affiliation{State Key Laboratory of Low Dimensional Quantum Physics, Department of Physics, Tsinghua University, Beijing 100084, China}
 \affiliation{Frontier Science Center for Quantum Information, Beijing 100193, China}

\date{\today}


\begin{abstract}
Time-reversed evolution has substantial implications in physics, including prominent applications in refocusing of classical waves\cite{Fink:2000aa, Lerosey:2004aa, Lerosey:2007aa} or spins\cite{Hahn:1950aa} and fundamental researches such as quantum information scrambling\cite{Shenker:2014aa, Garttner2017aa, Landsman:2019aa, Lewis-Swan:2019aa}. In quantum metrology\cite{Giovannetti:2004aa, Pezze2018}, nonlinear interferometry based on time reversal protocols\cite{Davis2016, Frowis2016,Macri:2016aa} supports entanglement-enhanced measurements without requiring low-noise detection. Despite the broad interest in time reversal, it remains challenging to reverse the quantum dynamics of an interacting many-body system as is typically realized by an (effective) sign-flip of the system's Hamiltonian. Here, we present an approach that is broadly applicable to cyclic systems for implementing nonlinear interferometry without invoking time reversal. Inspired by the observation that the time-reversed dynamics drives a system back to its starting point, we propose to accomplish the same by slaving the system to travel along a `closed-loop' instead of explicitly tracing back its antecedent path. Utilizing the quasi-periodic spin mixing dynamics in a three-mode $^{87}$Rb atom spinor condensate, we implement such a `closed-loop' nonlinear interferometer and achieve a metrological gain of $3.87_{-0.95}^{+0.91}$ decibels over the classical limit for a total of 26500 atoms. Our approach unlocks the high potential of nonlinear interferometry by allowing the dynamics to penetrate into deep nonlinear regime, which gives rise to highly entangled non-Gaussian state. The idea of bypassing time reversal may open up new opportunities in the experimental investigation of researches that are typically studied by using time reversal protocols.
\end{abstract}
\maketitle

%
%
%
Time reversal is an important concept in physics, supporting the understanding of the origin for `time's arrow'\cite{arrow} and applications in technologies such as time reversal mirrors\cite{Fink:2000aa, Lerosey:2004aa, Lerosey:2007aa} and spin or photon echos\cite{Hahn:1950aa, Kurnit:1964aa}.
Time reversal of quantum many-body dynamics is also of significant interest, due to its importance in investigating quantum information scrambling
\cite{Shenker:2014aa, Garttner2017aa, Landsman:2019aa, Lewis-Swan:2019aa}, diagnosing quantum phase transition or criticality\cite{Quan:2006aa, Nie:2020aa, Lewis-Swan:2020aa}, and developing entanglement-enhanced precision metrology\cite{Davis2016, Frowis2016,Macri:2016aa}. The commonly adopted approach for realizing time-reversed dynamics comes from time-forward evolution with a sign-flipped Hamiltonian. While simple and straightforward, this approach is generally difficult to realize in an interacting many-body system. Hence, developing approaches capable of bypassing the sign-flip of a Hamiltonian is of significant practical importance.

\begin{figure*}[t]
	\centering
	\includegraphics[width=0.95\linewidth]{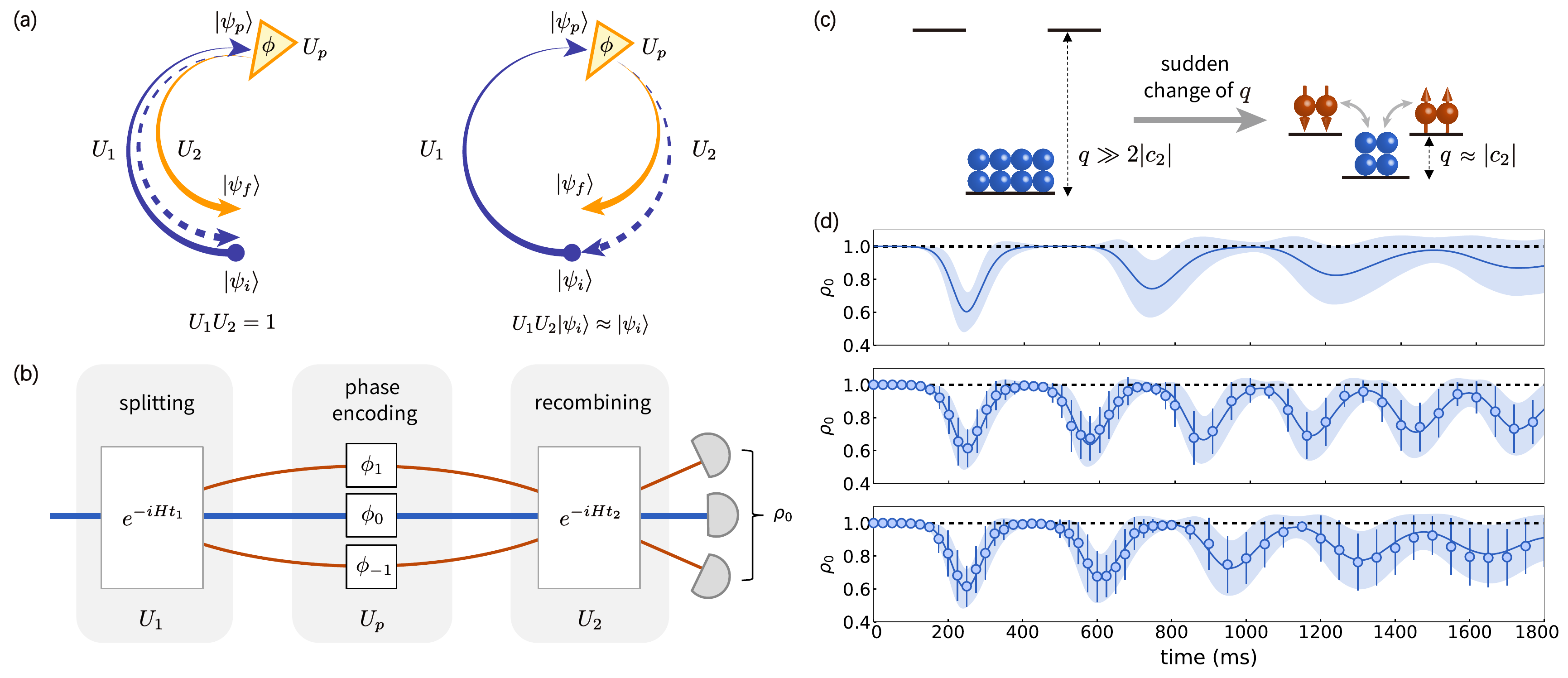}
	\caption{{\bf Three-mode nonlinear interferometry based on by cyclic dynamics.} (a) Nonlinear interferometry is typically realized by a pair of time-reversed operations: `path' splitting ${U}_1$ and recombining ${U}_2$, with phase encoding $U_p$ (orange triangle) sandwiched in between. The probe state $|\psi_p\rangle$ generated by ${U}_1$ returns back to the initial state $|\psi_i\rangle$ after ${U}_2$ if the phase encoded in between is $\phi=0$, while it ends up at a different state $|\psi_f\rangle$ if $\phi\neq0$. For a system with cyclic dynamics, the required time reversal can be circumvented by properly timing the instant to complete a periodic return.
(b) Schematic illustration for our three-mode nonlinear interferometer based on spin mixing dynamics (SMD) that creates (annihilates) paired atoms in $|1,\pm1\rangle$ components from (into) the $|1,0\rangle$ component. The corresponding SMD evolution effects nonlinear `path' splitting or recombining. The encoded relative phase $\phi=2\phi_0-(\phi_1+\phi_{-1})$ can be extracted from measured fractional population $\rho_0=N_0/N$ in the end.
(c) SMD is initiated by a sudden change of quadratic Zeeman shift (QZS) from $q\gg2|c_2|$ to $q\approx|c_2|$.
(d) Temporal evolution of $\rho_0$ starting with all atoms in the $|1,0\rangle$ component. The upper panel denotes the ideal case without atom loss or technical noises, where $q$ is set to be $0.99|c_2|$, while the middle and lower panels present experimentally measured data respectively for fixed $q=0.99|c_2|$ and tuned $q(t)=0.99|c_2|e^{-0.4\gamma t}$ with loss rate $\gamma=0.09~\rm{s^{-1}}$ (to compensate for slightly drifting $c_2$ due to decreasing $N$). All data points denote average over 30 experimental runs and error bars stand for 1 standard deviation. The solid
lines and shaded regions denote numerically simulated mean and standard deviation of $\rho_0$ based on the truncated Wigner method.
}
 \label{fig1}
\end{figure*}

In this study, we address the challenge of effecting time-reversed evolution in the context of quantum metrology\cite{Giovannetti:2004aa, Pezze2018}, aimed at beating
the standard quantum limit (SQL) by using entanglement.
Nonlinear interferometry based on time reversal protocol\cite{Davis2016, Frowis2016,Macri:2016aa}
was proposed to circumvent low-noise detection unanimously
required in entanglement-enhanced metrology based on
linear interferometry, where the improvement to measurement signal-to-noise ratio (SNR) comes from reduced quantum noise by correlations between entangled particles\cite{Gross2010, Riedel2010, Lucke2011, Hosten2016, Luo2017, Zou2018, Pedrozo-Penafiel:2020aa, Bao:2020aa}.
To benefit from such squeezed noise, however, other noises especially the readout (or detection) noise must be made smaller,
which is technically challenging for ensembles of large particle numbers\cite{Hume:2013aa, Qu:2020aa, Huper:2020aa}.
Nonlinear interferometry improves SNR by magnifying signal instead,
hence it is inherently robust against detection noise\cite{Davis2016, Frowis2016,Macri:2016aa}.
A typical nonlinear interferometer consists of three building blocks:
nonlinear `path' splitting (${U}_1$) for generating 
entangled probe state,
phase encoding (${U}_p$), and nonlinear `path' recombining (${U}_2$)
for transforming the encoded phase into easily measured observables.
Usually, the recombining is taken to be the time reversal of the splitting process, i.e., ${U}_2 = {U}_1^\dag$, such that the state traces back its entanglement generation trajectory and returns to the input state if no phase is encoded (see the left panel of Fig.~\ref{fig1}(a)).
The presence of a nonzero encoded phase, however, breaks such a closed-loop and
gives rise to a phase dependent output state. The capability of nonlinear interferometry for enhanced SNR has been demonstrated
in several pioneering experiments, ranging from photons\cite{Hudelist2014, Manceau2017} to Bose-Einstein condensate (BEC)\cite{Linnemann2016, Linnemann2017}, cold thermal atoms\cite{Hosten2016_mag}, and a mechanical oscillator\cite{Burd2019}.
As a cost of engineering time reversal, these experiments are typically constrained to short-term evolutions, where the effective time-reversed dynamics kicks in before the probe states become too deeply entangled to be disentangled, hence sacrificing potentially higher metrological gain.

Here we present a general approach for implementing
nonlinear interferometry without explicitly invoking time reversal.
The key idea is to employ cyclic dynamics, which automatically drives the system back to the vicinity of initial state, as a substitute for time reversal.
As shown in the right panel of Fig.~\ref{fig1}(a),
the complete interferometric protocol starts with
a classical product state $|\psi_i\rangle$. The subsequent evolution (clockwise)
under a many-body interaction Hamiltonian ${H}$ enacts
nonlinear splitting ${U}_1$ before the system arrives at an intermediate entangled probe state $|\psi_p\rangle$. In the absence of phase encoding, cyclic dynamics drives the system forward, clockwise towards the initial state after the second stage of complementary time-forward evolution ${U}_2$, which mimics the effect of time reversal of ${U}_1$ (shown by the counter-clockwise pointed dashed arc in the left panel of Fig.~\ref{fig1}(a)). Such an implementation not only circumvents the challenge of flipping the sign of Hamiltonian, it also enables evolution beyond the short-term limit, and enhances the metrological performance with highly entangled non-Gaussian probe states generated by long-term dynamics.

We demonstrate the above protocol in a $^{87}$Rb atom spinor BEC,
prepared initially in the $|F=1,m_F=0\rangle$ hyperfine ground state and 
described by the Hamiltonian\cite{Law1998},
\begin{equation}
H=\frac{c_2}{2N}\big[2({a}_1^{\dag}{a}_{-1}^{\dag}{a}_0{a}_0+\mathrm{h.c.})+(2{N}_0-1)(N-{N}_0)\big]-q{N}_0,
\end{equation}\label{Ham}
where ${a}_{m_F}^\dag$~(${a}_{m_F}$) denotes the creation~(annihilation) operator for the $m_F(=\pm1, 0)$ spin component,
${N}_{m_F}={a}_{m_F}^\dag{a}_{m_F}$ its atom number, and $N$ the total number of atoms.
The terms inside the square brackets describe spin mixing dynamics~(SMD) at a spin-exchange rate $|c_2|$, which creates (annihilates)
paired atoms in $|1,\pm1\rangle$ from (into) $|1,0\rangle$ components, as well as elastic collision caused energy shifts.
The last term describes an effective quadratic Zeeman shift~(QZS), tunable with magnetic field or off-resonance microwave. Linear Zeeman shift is omitted here due to the conservation of magnetization ($L_z=N_1-N_{-1}$).
\begin{figure*}[t]
	\centering
	\includegraphics[width=1.0\linewidth]{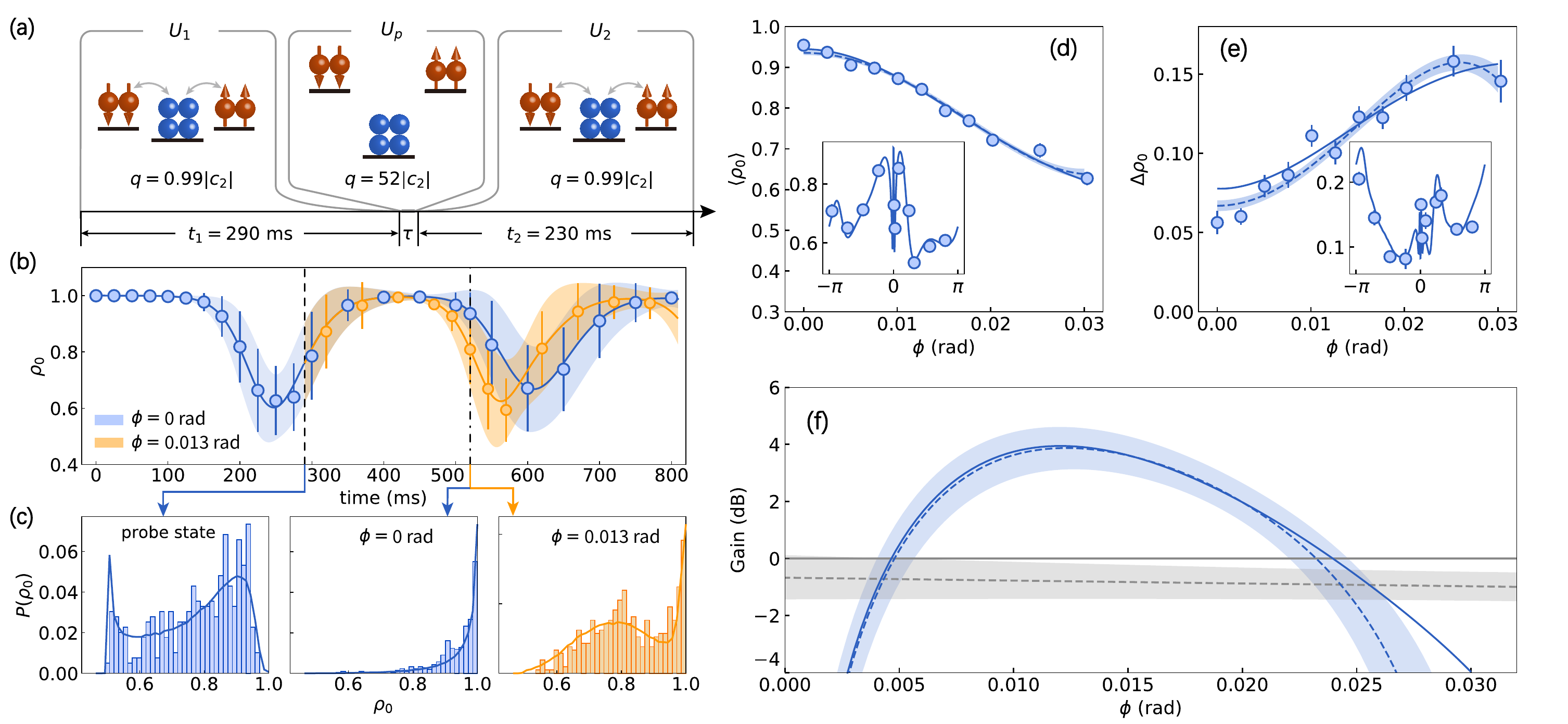}
\caption{{\bf Phase sensitivity.} (a) Our implementation of the nonlinear interferometry with `path' splitting and recombining both from time-forward SMD for $t_1=290~\rm{ms}$ and $t_2=230~\rm{ms}$ respectively. Spinor phase is encoded by quenching $q$ to $52|c_2|$ with microwave dressing and holding the condensate for variable time $\tau$. (b) Temporal evolution of $\rho_0$ for different $\phi$ encoded at the instant marked by vertical black dashed line. Error bars are the statistical uncertainty of over 30 experimental runs, and the shading denotes the simulated standard deviation. (c) Probability distributions of $\rho_0$ for the probe state and for the final states obtained from 400 experimental runs. (d) Mean value and (e)~standard deviation of $\rho_0$ in the vicinity of $\phi=0$ and within $[-\pi, \pi]$ (insets). The rapid variations of both quantities near $\phi\sim 0$ implicate our nonlinear interferometry is highly sensitive to small encoded phase. Each point represents the result of 100 experimental runs. (f) Metrological gain obtained from error propagation formula. We observe a highest gain of $3.87_{-0.95}^{+0.91}$~dB at $\phi=0.012$~rad. The blue dashed lines in (d-f) are biquadratic fitted curves to experimental data with the shaded regions indicating the fitting uncertainties. The grey dashed line (and shading) in (f) denotes the sensitivity (and its uncertainties) achieved in linear three-mode interferometry, while the grey solid line denotes the standard quantum limit (SQL) with respect to the total number of atoms. Solid blue and orange lines in (b-f) are the simulation results.}
\label{fig2}
\end{figure*}

In the undepleted pump regime with nearly all atoms in the $|1,0\rangle$ component, the operator ${a}_0$ can be approximated by a complex number $\sqrt{N}e^{i\phi_0}$, which reduces (1) to the widely discussed
SU(1,1) form\cite{Yurke1986} $(\kappa {a}_1^{\dagger}{a}_{-1}^{\dagger}+\kappa^*{a}_1{a}_{-1})$ that creates or annihilates paired atoms in the $|1,\pm1\rangle$ modes with strength $\kappa=c_2 e^{2i\phi_0}$ when QZS is adjusted to cancel the energy shift of elastic collsion at $q=|c_2|$.
The sign of this Hamiltonian can be flipped by imprinting a phase of $\pi/2$ on the $|1,0\rangle$ pump mode, which has been employed to realize the first atomic SU(1,1) interferometry\cite{Linnemann2016, Linnemann2017}.
While beating the classical limit with respect to the small number ($\sim 2.8$)
of atoms in the phase sensing modes
(namely the $|1,\pm1\rangle$ components), 
the precision realized was far below the SQL for the total atom number ($N\approx400$).

One can increase the number of atoms in the phase sensing modes to increase phase sensitivity, for example by extending SMD beyond the undepleted pump regime\cite{Gabbrielli2015}, or by linearly coupling the three spin components before phase sensing\cite{Szigeti2017}. This work
adopts the former strategy with the consequent interferometry based on the complete Hamiltonian (1) of
SMD
controlled by the relative strength of $q$ vs $|c_2|$, over extended time into the deep nonlinear regime. In a ferromagnetic system ($c_2<0$), the ground state at zero magnetization for $q\gg2|c_2|$ is a product state of all atoms occupying the $|1,0\rangle$ component (or a polar state). When suddenly quenched to $q\in (-2,2)|c_2|$, the initial polar state undergoes coherent many-body spin oscillation\cite{Chang2005}, whose dynamics can be mapped to that of a nonlinear pendulum in a semiclassical treatment\cite{Zhang2005, Gerving2012}.
Although the quantum superposition of unequally spaced energy eigenstates prevents exact pendulum-like periodic oscillations from evolving indefinitely, for a typical condensate with tens of thousands of atoms or more, the first period of collective state oscillation remains clearly recurrent\cite{Rauer2018aa,Schweigler:2021aa} nevertheless.
The cyclic dynamics then drives the system back to the immediate vicinity of the initial state, in line with the time-reversed evolution as discussed (see the upper panel of Fig.~\ref{fig1}(d)).
Based on this understanding, we construct a three-mode nonlinear interferometer as illustrated in Fig.~\ref{fig1}(b), with splitting and recombining effected by complementary time-forward SMD, for durations $t_1$ and $t_2$, respectively. With such a setup, as we will show below, the encoded relative phase $\phi=2\phi_0-(\phi_1+\phi_{-1})$ can be inferred at a precision below the classical limit with respect to the total number of atoms for all three spin components, by simply measuring the final population of $|1,0\rangle$.
%

Our experiments are carried out in an almost pure $^{87}$Rb BEC of $N\approx26,500$ atoms.
The bias magnetic field is fixed at $0.23~\rm{G}$ and stabilized by feedback control
to a temporal peak-to-peak fluctuation of $150~\rm{\mu G}$, 
corresponding to a QZS of $q_B=2\pi\times3.8~\rm{Hz}$ with a relative uncertainty of $0.001$.
Since this value of $q_B$ is well below the quantum critical point at $q_c=2|c_2|$
with $c_2=-2\pi\times3.85~\rm{Hz}$ calibrated experimentally\cite{Luo2017},
a microwave dressing field ($10~\rm{MHz}$ red detuned from the clock transition between
$|1,0\rangle$ and $|2,0\rangle$) is switched on during the preparation of initial state (see Methods), whose AC Stark shift augments the total QZS to $10|c_2|$
such that the condensate is maintained in the polar phase. To illustrate the near cyclic behavior, we switch off the microwave to quench the condensate to below the quantum critical point, and let the system freely evolve at $q=q_B=0.99|c_2|$ for $1.8~\rm{s}$. At the end of the evolution, the trap is turned off and atoms in different spin components are spatially separated using a Stern-Gerlach pulse followed by $10~\rm{ms}$ time-of-flight expansion, and are finally detected with low-noise absorption imaging\cite{Luo2017}.

The system's evolution in terms of the
measured fractional population $\rho_0=N_0/N$ is shown in the middle panel of Fig.~\ref{fig1}(d).
Compared to the expected dynamics free from external noise or loss (upper panel), we find a clear deviation starting from the end of the first oscillation period. The prominent plateau, or the collapsed region between the first two oscillation troughs is shortened, and the subsequent oscillation features larger frequency and amplitude. Such seemingly improved coherence in fact arises from decoherence due to mechanisms including particle loss\cite{Gerving2012} and weak radio-frequency (RF) noise from electronic devices around the BEC chamber. Taking these imperfections into considerations, we find the measured data agrees well with the numerical simulations based on the truncated Wigner method (see Supplementary Information for more details). The effect of loss can be partially mitigated by tuning the bias $B$-field to maintain a fixed ratio of $q/|c_2|$, which compensates for the drift of $c_2\propto N^{2/5}$ from decreasing $N$, and leads to a better agreement of the data with ideal dynamics (slower oscillation as well as faster damping, shown in the lower panel of Fig.~\ref{fig1}(d)). Such a compensation improves the metrological performance of our system, and is therefore adopted in the reported interferometry.

The specific sequence of the implemented nonlinear interferometry protocol is illustrated in Fig.~\ref{fig2}(a).
We first hold the condensate (initially in the polar phase) for $t_1=290~\rm{ms}$ (marked by the vertical black dashed line in Fig.~\ref{fig2}(b)) at $q/|c_2|=0.99$.
At this instant, the system is already far beyond the undepleted pump regime, and the corresponding generated state takes on a highly non-Gaussian distribution (Fig.~\ref{fig2}(c) left panel) with around $22\%$ of the atoms transferred to $|1,\pm1\rangle$ components.
For phase sensing, we switch on the dressing microwave for a variable time $\tau$, which quenches $q$ to $52|c_2|$ to sufficiently suppress SMD, and imprints a phase $\phi=2q\tau$ on the condensate via $U_p=e^{i\phi  N_0/2}$ (see Methods).
Finally, the interferometer is completed by resuming SMD for a second time-forward duration $t_2=230~\rm{ms}$
with the microwave field switched off.
\begin{figure}[t]
	\centering
	\includegraphics[width=1\linewidth]{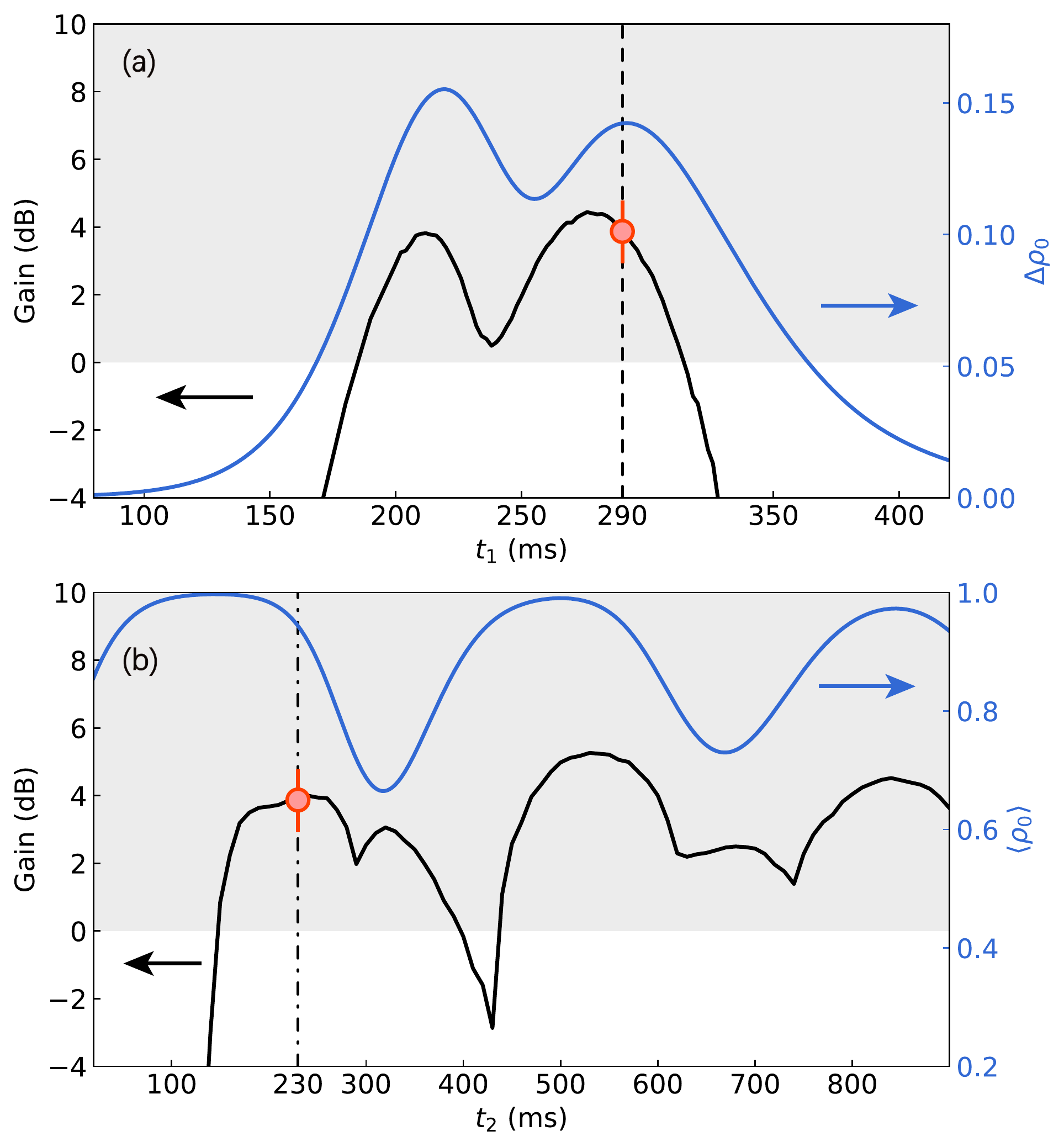}
\caption{{\bf The dependence of metrological gain on spin mixing times $t_1$
and $t_2$}. (a) The `path' splitting time $t_1$ is scanned while the total evolution
time $t_1 + t_2$ is fixed at $520~\rm{ms}$. The gains optimized over $\phi\in\left[0, 0.03~\rm{rad}\right]$
for every point share similar dependence on $t_1$ as $\Delta\rho_0$. (b) For $t_1$
fixed at $290~\rm{ms}$, the phase-optimized gain oscillates with $t_2$, almost in
sync with $\langle\rho_0\rangle$. The local maxima appear at $\langle\rho_0\rangle\approx1$, when `path' recombining
nearly reverses `path' splitting. Black dashed and dash-dotted vertical lines denote the durations
employed in our experiments. Red circles represent the experimental data point, and grey shaded regions indicate metrological gain beyond SQL.}
\label{fig3}
\end{figure}

Figure~\ref{fig2}(b) presents time evolution of measured $\rho_0$ 
for small but different phase $\phi$.
In the absence of an encoded phase, the system nearly returns to its initial state at the end of the interferometry (marked by the vertical dash-dotted line), with the probability distribution $P(\rho_0)$ concentrated around $\rho_0=1$ (middle panel of Fig.~\ref{fig2}(c)).
A nonzero phase shift of $2/\sqrt{N}\approx0.013~\rm{rad}$ causes the second oscillation cycle to inch forward,
leading to an evident decrease in the mean value of $\rho_0$ as well as broadening of its distribution (right panel in Fig.~\ref{fig2}(c)). The long tail of this distribution is near-Gaussian, which enables us to extract the encoded phase (with a high sensitivity) by measuring the mean value of $\rho_0$.
The phase sensitivity shown in Fig.~\ref{fig2}(f) is obtained from error propagation formula~$\Delta\phi={\Delta\rho_0}/{|d\langle\rho_0\rangle/d\phi|}$, where the denominator and the numerator are obtained from fitting experimental data to biquadratic functions (blue dashed lines in Fig.~\ref{fig2}(d-f)).
We define the metrological gain as $G=-20\log_{10}(\Delta\phi/\Delta\phi_{\rm{SQL}})$ with respect to the SQL $\Delta\phi_{\rm{SQL}}=2/\sqrt{N}$, the optimal phase sensitivity achievable for a coherent spin state (CSS) in a three-mode linear interferometer with the same phase imprinting operator (Supplementary Information).
To benchmark this reference value, we prepare a CSS with
single-particle wavefunction $|\psi\rangle=-\frac{i}{2}|1,1\rangle+\frac{1}{\sqrt{2}}|1,0\rangle-\frac{i}{2}|1,-1\rangle$,
which is known theoretically to saturate the SQL,
and then measure the angular momentum $L_x$ after phase encoding (see Methods). The obtained phase sensitivity is indeed found to be around SQL as shown by the grey dashed line in Fig.~\ref{fig2}(f).
In comparison, for our nonlinear interferometer, we observe a maximal gain of $3.87_{-0.95}^{+0.91}$ dB beyond this SQL at $\phi=0.012~\rm{rad}$.


The metrological performance of the implemented nonlinear interferometer crucially depends on the spin mixing times $t_1$ and $t_2$.
The `path' splitting part of SMD for $t_1$ determines the probe state $|\psi_p\rangle$ used for phase encoding, whose multi-particle entanglement is ultimately responsible for
observing phase sensitivity beyond SQL.
The highest sensitivity $\Delta\phi =1/\sqrt{{\cal F}_Q}$ achievable is characterized by quantum Fisher information ${\cal F}_Q$ of the probe state\cite{Pezze2018}. Thus ideally, one should employ a probe state with an ${\cal F}_Q$ as large as possible in order to optimize the interferometric gain.
For phase encoding with the generator ${N}_0/2$ employed here, ${\cal F}_Q=(\Delta  N_0)^2=(N\Delta \rho_0)^2$ in the absence of decoherence, which equals to the variance of population in the $|1,0\rangle$ component.

The blue line in Fig.~\ref{fig3}(a) shows the simulated $\Delta \rho_0$ as a function of $t_1$. The interferometer therefore is expected to perform well in the vicinity of $t_1$, where $\Delta \rho_0$ reaches a peak\cite{qfi}. This is indeed what we observe from the numerical simulations (including particle loss and RF noise), where we fix the total time $t_1 + t_2$ and investigate the dependence of the optimal phase sensitivity on $t_1$ (black solid curve). 
In the experiment, we work at $t_1 = 290$ ms, as marked by the vertical dashed line.
The probe state generated at this instant is highly non-Gaussian, as shown in the left panel of Fig.~\ref{fig2}(c). Although such a state is capable of providing a high phase sensitivity, it is difficult to reach this by using linear interferometry, where the output state remains non-Gaussian and thus the measurement of high order moments\cite{Gessner2019} or even full probability distribution\cite{Strobel2014} will be required. In contrast, the nonlinear interferometry protocol reported here gives an output state with a nearly Gaussian distribution, which makes it possible to obtain a high phase sensitivity based only on the mean value and standard deviation of $\rho_0$. Our work therefore demonstrates an implementable method to
exploit highly entangled non-Gaussian states for quantum metrology, which until now are rarely utilized due to the associated complexity of characterizations\cite{Strobel2014}.


Next we investigate the recombining part of the nonlinear interferometer. In Fig.~\ref{fig3}(b), we keep splitting time $t_1$ (therefore the probe state) fixed while scan $t_2$. Phase sensitivity beyond SQL is found over a wide range, and the metrological gain oscillates with $t_2$, almost in sync with $\langle\rho_0\rangle$. Especially worthy of pointing out is the fact that local maxima of gain appear near the maxima of $\langle\rho_0\rangle$. At these moments, the state returns back to the close vicinity of the initial polar state, or in other words, the `path' recombining most closely resembles the time reversal of the splitting.
This further confirms the feasibility of our protocol for bypassing time reversal.
It is also noted that there exists a short delay between the first maxima of gain and that of $\langle\rho_0\rangle$, which is attributed to our specific characterization of final state through measuring only the mean value and standard deviation of $\rho_0$. This delay disappears when the full distribution of $\rho_0$ is used, as shown in the Supplementary Information.

The nonlinear interferometer we implement is robust to detection noise. This can be appreciated by noting that detection noise, around 20 (atoms) in our system, is almost two orders of magnitude smaller than the measured number fluctuation of atoms in $|1,0\rangle$ components (see Fig.~\ref{fig2}(e)).
The achievable phase sensitivity is currently limited by technical imperfections including atom loss and RF noise due to the long evolution time required to complete the `closed-loop'.
This loop, as specifically implemented in our work, is based on cyclic dynamics, while more flexible approaches can be actively sought for by using optimal control\cite{walmsley2003quantum} or machine learning\cite{Carleo:2019aa, Guo:2021aa}.
Our idea for bypassing time reversal may open up new opportunities in the experimental investigation of researches that are typically studied by using time reversal protocols.

\bibliography{ref}

~\\
\textbf{Data availability} All data that support the plots within this paper and other findings of this study are available from the corresponding author upon reasonable request.\par
~\\
\textbf{Code availability}
All relevant codes or algorithms are available from the corresponding author upon reasonable request.\par
~\\
\textbf{Acknowledgements}
We thank F. Chen, Y. Q. Zou, J. L. Yu and M. Xue for helpful discussions.
This work is supported by the National Natural Science
Foundation of China (NSFC) (Grants  No. 11654001, No. U1930201, No. 91636213, and No. 91836302), by the Key-Area Research and Development Program of GuangDong Province
(Grant No. 2019B030330001), and by the National Key R\&D Program of China (Grants No. 2018YFA0306504 and No. 2018YFA0306503).\par
~\\
\textbf{Author contributions}
L.-N.W. and L.Y. conceived the study. Q.L., J.-H.C., T.-W.M. and S.-F.G. performed the experiment and analysed the data. Q.L., L.-N.W. and X.-W.L. conducted the numerical simulations. Q.L., L.-N.W., M.K.T. and L.Y. wrote the paper.\par
~\\
\textbf{Competing interests} The authors declare no competing interests.\par

\clearpage

\section*{Methods}
\setcounter{figure}{0}
\renewcommand{\figurename}{Extended Data Fig.}
\begin{figure}[b]
	\centering
	\includegraphics[width=1\linewidth]{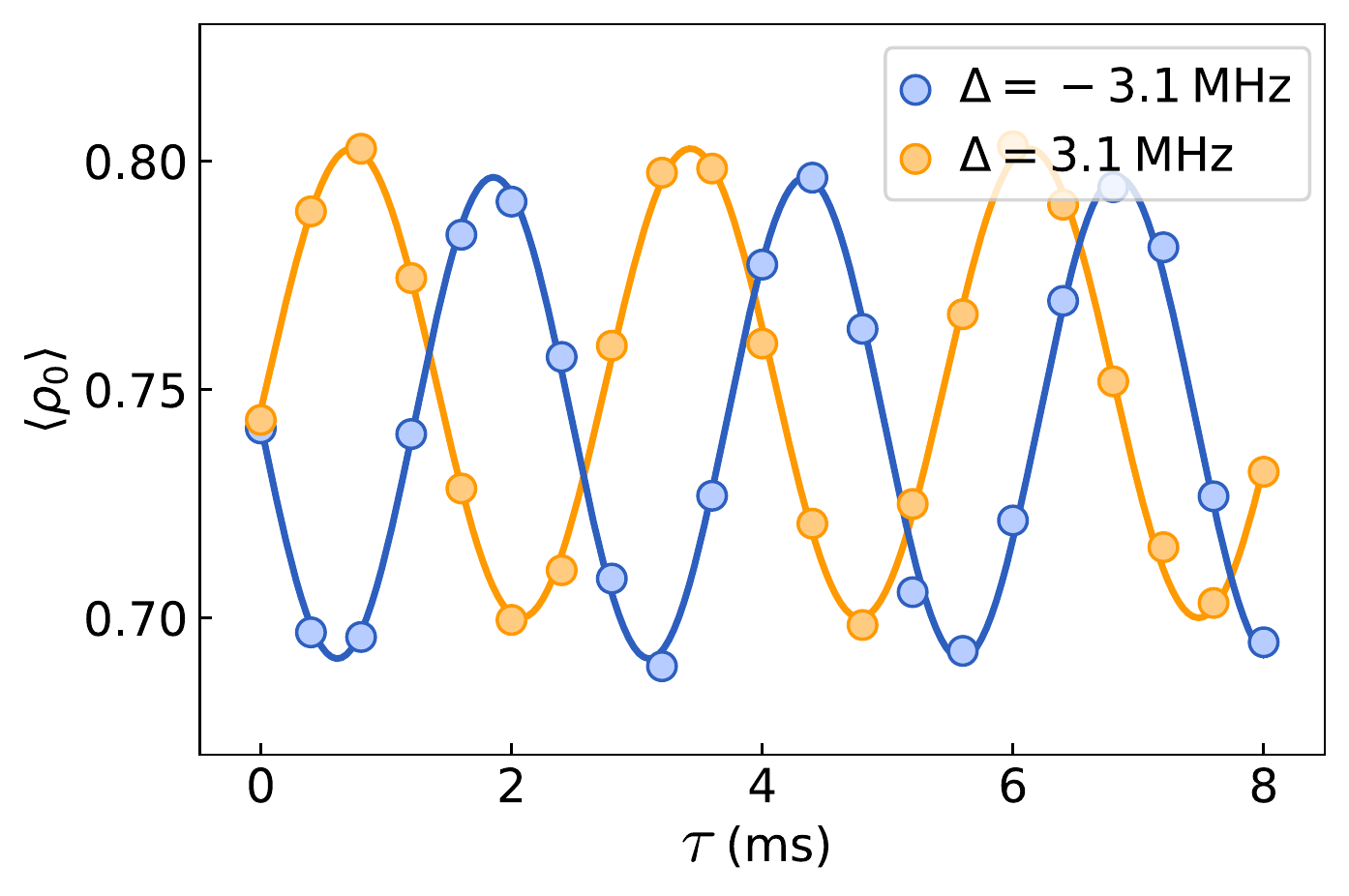}
	\caption{Calibration of QZS during phase imprinting. For $q_t\gg|c_2|$, the state will accumulate a time dependent relative phase without undergoing noticeable population changes in three Zeeman sublevels. Following the short-term spin dynamics for $5$ ms at $q\approx|c_2|$, the imprinted phase converts to fringes of $\langle\rho_0\rangle$ oscillation at a frequency $2q_t$. Blue (orange) dots denote data points measured with microwave detuning of $\Delta=-3.1 (3.1)~\rm{MHz}$ for the `dressing' (`pre-dressing') pulse. Fitting them with sinusoidal function Eq.~\eqref{fitq} enables us to determine the value of $q_t$. Solid lines refer to the corresponding fitting curves.}
	\label{cali_q}
\end{figure}
\subsubsection*{Initial state preparation}\label{prep}
We prepare a $^{87}$Rb BEC of around 27500 atoms in $|1, 0\rangle$ hyperfine ground state, confined by a crossed optical dipole trap with trapping frequencies $2\pi\times(209, 97, 169)$ Hz. The initial bias magnetic field is $0.8$ G, which gives a QZS of $2\pi\times46~\rm{Hz}$. To initiate SMD, we need to compensate for such a large QZS with microwave dressing. However, as the relative stability of the microwave power is about $0.001$, direct dressing will lead to a fluctuation of QZS on the order of $0.01|c_2|$, which could severely degrade the coherence of spin dynamics. Instead, we first lower the magnetic field to $0.23$ G within $200$ ms. The resulting QZS matches $|c_2|$ in the absence of microwave dressing, while has a much smaller peak-to-peak fluctuation of $0.001|c_2|$ arising from the fluctuating magnetic field. To inhibit spin mixing during the ramping process, we switch on microwave field $10$ MHz red-detuned from $|1,0\rangle$ to $|2,0\rangle$ transition to keep QZS above $10|c_2|$. After the ramping, the condensate is hold for another $100$ ms to assure the magnetic field is sufficiently stabilized. During the ramping process, ambient RF noise may transfer a tiny amount of atoms from $|1,0\rangle$ to $|1,\pm1\rangle$, which can be carefully removed by two consecutive resonant microwave $\pi$ pulses to transfer atoms in the $|1,\pm1\rangle$ to $|2,\pm2\rangle$ states, and cleaned out with a flush of resonant probe beam. The total atom number is around 26500 after this operation. The SMD is then initiated by switching off the dressing microwave field.

\subsubsection*{Phase imprinting}\label{imprint}
Spinor phase can be imprinted by shifting the QZS with microwave dressing.
To sufficiently halt the SMD during phase imprinting, we switch on the microwave field $3.1$ MHz red-detuned from the $|1,0\rangle$ to $|2,0\rangle$ transition to a power of $0.6$ W. The resulting QZS reaches $q=2\pi\times200.9(6)$ Hz, which corresponds to $52|c_2|$, much higher than the quantum critical point $q_c=2|c_2|$. The power of microwave pulse (denoted as `dressing' pulse) is linearly ramped up and down within $12~\rm{\mu s}$ respectively to avoid sideband excitation. Measurement of phase sensitivity requires a precise control of small imprinted phase in the vicinity of $\phi=0$, which is realized by using composite pulses as adopted in our earlier work\cite{Zou2018}. Specifically, we apply a `pre-dressing' microwave pulse with opposite detuning, and optimize its amplitude to cancel the phase shift induced by the rising and trailing edges of the `dressing' pulse. The net accumulated phase is then given by $\phi=2q\tau$, where $\tau$ denotes the duration of the `dressing' pulse without the edges.

\subsubsection*{Calibration of QZS}\label{cali}
Starting from a polar state, we apply a resonant RF Rabi pulse to transfer a quarter of the atoms to $|1,\pm1\rangle$ modes. The microwave field is then switched on, which quenches QZS to the target value $q_t$ for measurement. For $q_t\gg|c_2|$, the SMD is energetically suppressed, therefore the subsequent evolution does not lead to noticeable population changes in the Zeeman sublevels, but provides a time dependent relative phase $\phi(\tau)=2q_t\tau$ instead. After holding the condensate for variable time, we switch off the microwave to initiate SMD at $q=|c_2|$ and let the system evolve for $t=5$ ms. At the end of such a short-term evolution, the change of $\langle\rho_0\rangle$ can be approximated as
\begin{equation}\label{fitq}
\langle\rho_0(\tau)\rangle-\rho_0\approx2c_2t\rho_0(1-\rho_0)\sin(2q_t\tau),
\end{equation}
where $\rho_0\approx0.75$ denotes the value before evolution, from linearizing the mean field differential equation\cite{Zhang2005} at $t=0$. Therefore, by tuning the phase accumulation time $\tau$ and fitting the final value of $\langle\rho_0\rangle$ with sinusoidal function, we calibrate the QZS to be $2\pi\times200.9(6)$ Hz and $-2\pi\times185.0(4)$ Hz for the `dressing' and `pre-dressing' microwave pulses respectively~(as shown in Extended Data Fig.~1).

\begin{figure}[t]
	\centering
	\includegraphics[width=1\linewidth]{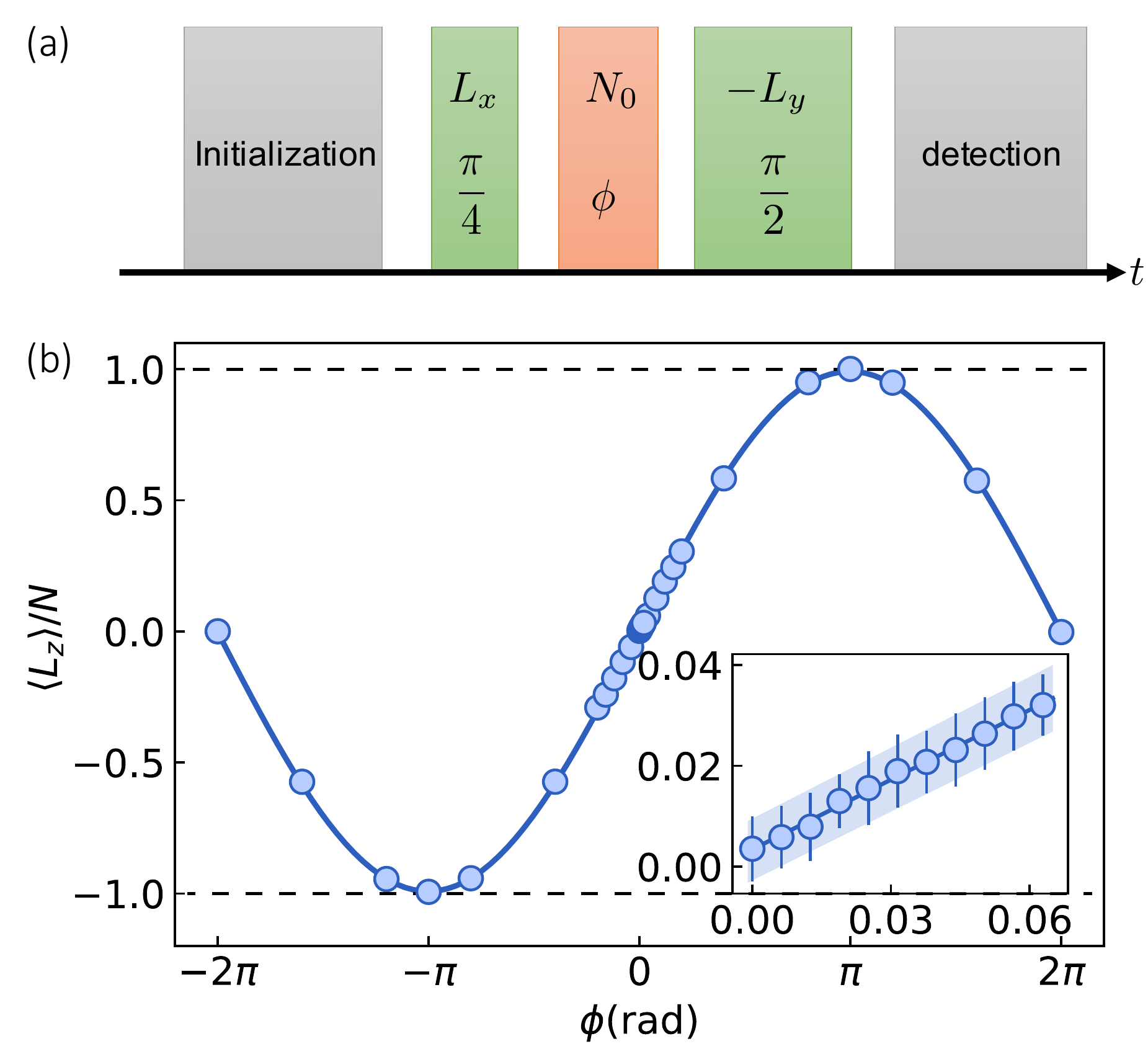}
	\caption{(a) Schematic illustration of the experimental sequence for the three-mode linear interferometer. (b) Data in the main figure shows single-shot results, while the solid line is the fitted curve based on sinusoidal function. Two horizontal black dashed lines denote $\langle L_z\rangle/N=\pm1$ respectively. In the inset, we obtain the mean value and fluctuation in the vicinity of $\phi=0$ by repeating the measurement 50 times for every point. Error bars denote one standard deviation and the breadth of shaded area refers to the quantum projection noise for ideal coherent spin state. Solid line in the inset denotes the result of linear fit.}
	\label{fig_linear}
\end{figure}

\subsubsection*{Three-mode Linear Interferometer}\label{linear}
To benchmark the SQL, we implement a three-mode linear interferometer. With polar state as input, a $\pi/4$ RF pulse rotates it around $L_x$ axis, and gives rise to the single particle state $|\psi\rangle=-\frac{i}{2}|1,1\rangle+\frac{1}{\sqrt{2}}|1,0\rangle-\frac{i}{2}|1,-1\rangle$ for phase sensing. Since $\langle\psi|\rho_0|\psi\rangle=0.5$, this state provides the optimal phase sensitivity among all coherent spin states (see Supplementary Information for the discussion of SQL). Accumulation of phase $\phi$ under the action of $ N_0$ will lead to phase-dependent mean value and standard deviation of $ L_x$. A final RF $\pi/2$ pulse along $-L_y$ axis converts the phase signal to population imbalance in $|1,\pm1\rangle$ components, leading to $\langle L_z\rangle=N\sin(\phi/2)$ and $\Delta L_z=\sqrt{N}|\cos(\phi/2)|$ for $N$ atoms~(see Extended Data Fig.~2). Error propagation formula then gives the phase sensitivity as $\Delta\phi=2/\sqrt{N}$, which coincides with the SQL we defined.\par

The linear interferometer is implemented at a bias magnetic field of $0.4$ G. The first RF pulse takes $10~\rm{\mu s}$, corresponding to a Rabi frequency of $2\pi\times12.5$ kHz. The phase is imprinted by applying two resonant $\pi$ pulses coupling $|1,0\rangle$ and $|2,0\rangle$. Atoms in $|1,0\rangle$ component accumulate a geometric phase afterwards, which can be flexibly adjusted by tuning the relative phase of the microwave pulses\cite{Hamley2012}. Each of the pulses takes $12~\rm{\mu s}$, and their amplitudes follow the Blackman profile in order to reduce crosstalk among other spin levels. Ideally, the phase imprinting process should only change the spinor phase ($\phi=2\phi_0-\phi_1-\phi_{-1}$). However, the microwave pulses will also induce unbalanced AC stark shifts to $|1,\pm1\rangle$ components, thereby generate an extra change to the Larmor phase ($\phi_l=\phi_1-\phi_{-1}$). Fortunately, since this latter phase only depends on the power and frequency of the applied microwave pulses, we can treat it as a constant offset and compensate for it by tuning the phase of the following RF pulse. An additional RF spin echo pulse is applied in the middle of the interferometer to mitigate the decoherence due to slow drift of magnetic field.\par





\end{document}


\title{Supplemental material: Nonlinear interferometry beyond classical limit facilitated by cyclic dynamics}

\author{Qi Liu*}
\affiliation{State Key Laboratory of Low Dimensional Quantum Physics, Department of Physics, Tsinghua University, Beijing 100084, China}
\author{Ling-Na Wu*}
\affiliation{Institut f\"ur Theoretische Physik, Technische Universit\"at Berlin,\\
Hardenbergstra\ss e 36, Berlin 10623, Germany}

\author{Jia-Hao Cao}
\affiliation{State Key Laboratory of Low Dimensional Quantum Physics, Department of Physics, Tsinghua University, Beijing 100084, China}
\author{Tian-Wei Mao}
\affiliation{State Key Laboratory of Low Dimensional Quantum Physics, Department of Physics, Tsinghua University, Beijing 100084, China}
 \author{Xin-Wei Li}
\affiliation{State Key Laboratory of Low Dimensional Quantum Physics, Department of Physics, Tsinghua University, Beijing 100084, China}
\author{Shuai-Feng Guo}
\affiliation{State Key Laboratory of Low Dimensional Quantum Physics, Department of Physics, Tsinghua University, Beijing 100084, China}
\author{Meng Khoon Tey}
 \affiliation{State Key Laboratory of Low Dimensional Quantum Physics, Department of Physics, Tsinghua University, Beijing 100084, China}
 \affiliation{Frontier Science Center for Quantum Information, Beijing 100193, China}
\author{Li You}
 \affiliation{State Key Laboratory of Low Dimensional Quantum Physics, Department of Physics, Tsinghua University, Beijing 100084, China}
 \affiliation{Frontier Science Center for Quantum Information, Beijing 100193, China}

\date{\today}
\maketitle
This supplementary material contains related details not included in the main text. It documents (i) simulation methods of unitary and dissipative spin mixing dynamics in Sec.~\ref{simu} and (ii) definition of standard quantum limit and phase estimation methods in Sec.~\ref{sens}.
\section{Simulation of Spin Mixing Dynamics}\label{simu}
\subsection{Unitary Evolution}
For spin-1 $^{87}$Rb atom Bose-Einstein condensates (BEC) tightly confined in an optical trap, its dynamics can be simulated under the single-mode approximation (SMA)~\cite{Law:1998aa}, which assumes the same spatial wave function for all the spin components and only the spin degree of freedoms evolves. The system is then described by the following Hamiltonian
\begin{equation}
\begin{aligned}
{ H} = &\dfrac{c_2}{2N}\big[2({ a}_1^{\dag}{ a}_{-1}^{\dag}{ a}_0{ a}_0+{ a}_1{ a}_{-1}{ a}_0^{\dagger}{ a}_0^{\dagger})+({N}_1-{N}_{-1})^2\\
&+(2{ N}_0-1)({ N}_1+{N}_{-1})\big]-q { N}_0.
\end{aligned}
\label{ham}
\end{equation}
Here ${a}_i$ (${a}_i^{\dag}$) is the annihilation (creation) operator for atoms in the $|F=1, m_F=i\rangle$~($i=\pm1,0$) spin component and ${ N}_i$ is the associated number operator. The c-number parameters $c_2$, $q$, and $N$ respectively denote the spin-exchange interaction strength, quadratic Zeeman shift (QZS), and total atom number. Numerical simulation can be implemented by directly diagonalizing the Hamiltonian within the Fock state basis $|N_1, N_0, N_{-1}\rangle$. Due to the conservation of total magnetization ($ L_z= N_1- N_{-1}$), for dynamics starting from polar state ($|\psi(0)\rangle=|0,N,0\rangle$) as considered in this work, it is sufficient to work in the $L_z=0$ subspace spanned by $|k, N-2k, k\rangle(0\leq k\leq\lfloor N/2\rfloor)$. The corresponding Hamiltonian is a tridiagonal sparse matrix with elements
\begin{equation}
H_{k,k'}=\left\{
\begin{aligned}
& 0, |k-k'|>1, \\
& \dfrac{c_2}{N}\bigg[(N-2k)-\dfrac{1}{2}\bigg]2k-q(N-2k), k=k',\\
& \dfrac{c_2}{N}\sqrt{(N-2k)(N-2k-1)}(k + 1), k=k'-1,\\
& \dfrac{c_2}{N}\sqrt{(N-2k+1)(N-2k+2)}k, k=k'+1.
\end{aligned}
\right.
\end{equation}
The diagonalization of the Hamiltonian gives the matrix $V$ composed of eigenstates and $D$ for eigenvalues, which are related through $HV=VD$. The instantaneous wavefunction is obtained by
\begin{align}
|\psi(t)\rangle&=e^{-i Ht}|\psi(0)\rangle=Ve^{-iDt}V^{-1}|\psi(0)\rangle.
\label{quant_evolve}
\end{align}

\subsection{Evolution with Dissipation}

To properly simulate the experiments, it is necessary to take atom loss into account.
Spin mixing dynamics in the presence of dissipation can be described by the master equation
\begin{equation}
\begin{aligned}
\frac{d{\rho}}{dt}=-i[{H}(t),{\rho}]+&\sum_{i=0,\pm1}\gamma_i {\cal D}[{a}_i]{\rho},
\end{aligned}
\label{master}
\end{equation}
with ${\cal D}[{a}_i]{\rho}={a}_i{\rho}{a}_i^{\dagger}-\{{a}_i^{\dagger}{a}_i,{\rho}\}/2$ for dissipative rate $\gamma_i$.
Here, the atom loss is approximately modeled by single-particle dissipative channel, although in reality the initial loss is dominated by three-body recombination~\cite{Soding:1999aa}. The loss rate $\gamma_i=\gamma$ is assumed the same for all spin components, and ${H}(t)$ is the time-dependent Hamiltonian obtained by replacing the $N$ and $c_2$ in Eq.~\eqref{ham} by $N(t)=N e^{-\gamma t}$ and $c_2(t)=c_2 e^{-2\gamma t/5}$, where the prefactor $2/5$ in the exponent comes from the dependence on $N$ in the Thomas-Fermi approximation.
In principle, the above master equation can be numerically solved with quantum Monte-Carlo wave function method~\cite{Molmer:1993aa}. However, the calculation for large atom number becomes time-consuming. As an alternative, we resort to the semiclassical truncated Wigner method (TWM)~\cite{Blakie:2008aa}.
The mapping from master equation to the equation of motion for the Wigner function can be realized by making use of the following substitution rules:
\clearpage
\begin{widetext}
\begin{equation}
\begin{aligned}
&{a}_i\rho\leftrightarrow\left(\psi_i+\dfrac{1}{2}\dfrac{\partial}{\partial\psi_i^*}\right)W,\quad{a}_i^{\dagger}\rho\leftrightarrow\left(\psi_i^*-\dfrac{1}{2}\dfrac{\partial}{\partial\psi_i}\right)W,\\
&\rho{a}_i\leftrightarrow\left(\psi_i-\dfrac{1}{2}\dfrac{\partial}{\partial\psi_i^*}\right)W,\quad\rho{a}_i^{\dagger}\leftrightarrow\left(\psi_i^*+\dfrac{1}{2}\dfrac{\partial}{\partial\psi_i}\right)W,\\
\end{aligned}
\label{rules}
\end{equation}
where $W$ is the Wigner function with complex c-numbers $\psi_i(\psi_i^*)(i=\pm1,0)$.
Applying the above rules to Eq.~\eqref{master}, and neglecting the third-order derivatives, we obtain the corresponding equation
\begin{equation}
\begin{aligned}
\dfrac{\partial W}{\partial t}&=\bigg\{\dfrac{\partial}{\partial\psi_1}\big\{ic_2'\left[\psi_0^2\psi_{-1}^*+(|\psi_1|^2-|\psi_{-1}|^2+|\psi_0|^2)\psi_1\right]+\dfrac{\gamma}{2}\psi_1\big\}+\\
&\dfrac{\partial}{\partial\psi_0}\big\{ic_2'\left[2\psi_1\psi_{-1}\psi_0^*+\left(|\psi_1|^2+|\psi_{-1}|^2\right)\psi_0\right]-iq\psi_0+\dfrac{\gamma}{2}\psi_0\big\}+\\
&\dfrac{\partial}{\partial\psi_{-1}}\big\{ic_2'\left[\psi_0^2\psi_1^*+(|\psi_{-1}|^2-|\psi_1|^2+|\psi_0|^2)\psi_{-1}\right]+\dfrac{\gamma}{2}\psi_{-1}\big\}+\\
&\dfrac{\gamma}{4}\dfrac{\partial^2}{\partial\psi_1\partial\psi_1^*}+\dfrac{\gamma}{4}\dfrac{\partial^2}{\partial\psi_0\partial\psi_0^*}+\dfrac{\gamma}{4}\dfrac{\partial^2}{\partial\psi_{-1}\partial\psi_{-1}^*}+h.c.
\bigg\}W,
\end{aligned}
\end{equation}
%
with both drift and diffusion terms easily identified. Following the textbook transformation~\cite{book}, we can unravel it into coupled Ito stochastic differential equations (SDEs) as given in the following:
\begin{equation}
\left\{
\begin{aligned}\label{sde}
d\psi_1 &=-ic_2'\left[\psi_0^2\psi_{-1}^*+(|\psi_1|^2-|\psi_{-1}|^2+|\psi_0|^2)\psi_1\right]dt-\frac{\gamma}{2}\psi_1dt+\sqrt{\frac{\gamma}{2}}d\xi_1(t),\\
d\psi_0 &=-ic_2'\left[2\psi_1\psi_{-1}\psi_0^*+\left(|\psi_1|^2+|\psi_{-1}|^2\right)\psi_0\right]dt+iq\psi_0dt-\frac{\gamma}{2}\psi_0dt+\sqrt{\frac{\gamma}{2}}d\xi_0(t),\\
d\psi_{-1} &=-ic_2'\left[\psi_0^2\psi_1^*+(|\psi_{-1}|^2-|\psi_1|^2+|\psi_0|^2)\psi_{-1}\right]dt-\frac{\gamma}{2}\psi_{-1}dt+\sqrt{\frac{\gamma}{2}}d\xi_{-1}(t).
\end{aligned}
\right.
\end{equation}
Here $c_2^\prime={c_2(t)}/{N(t)}$ and $d\xi_i(t)$ are complex Wiener noise increment satisfying $\overline{d\xi_i(t)}=0$ and $\overline{d\xi_i^*(t)d\xi_j(t)}=\delta_{i,j}dt$.\par
In addition to atom loss, we find the spin dynamics is also susceptible to radio-frequency (RF) magnetic field noise in the transverse plane, which couples the three spin components and therefore slightly alters their populations. To simulate this effect on spin dynamics, we include a fluctuating term $\omega(t){L}_x$ into the Hamiltonian Eq.~\eqref{ham}. The SDEs then become
\begin{equation}
\left\{
\begin{aligned}\label{rfsde}
d\psi_1 &=-ic_2'\left[\psi_0^2\psi_{-1}^*+(|\psi_1|^2-|\psi_{-1}|^2+|\psi_0|^2)\psi_1\right]dt-\frac{\gamma}{2}\psi_1dt+\sqrt{\frac{\gamma}{2}}d\xi_1(t)-\dfrac{i}{\sqrt{2}}d\chi(t)\psi_0,\\
d\psi_0 &=-ic_2'\left[2\psi_1\psi_{-1}\psi_0^*+\left(|\psi_1|^2+|\psi_{-1}|^2\right)\psi_0\right]dt+iq\psi_0dt-\frac{\gamma}{2}\psi_0dt+\sqrt{\frac{\gamma}{2}}d\xi_0(t)-\dfrac{i}{\sqrt{2}}d\chi(t)\left(\psi_1+\psi_{-1}\right),\\
d\psi_{-1} &=-ic_2'\left[\psi_0^2\psi_1^*+(|\psi_{-1}|^2-|\psi_1|^2+|\psi_0|^2)\psi_{-1}\right]dt-\frac{\gamma}{2}\psi_{-1}dt+\sqrt{\frac{\gamma}{2}}d\xi_{-1}(t)-\dfrac{i}{\sqrt{2}}d\chi(t)\psi_0,
\end{aligned}
\right.
\end{equation}
where $d\chi(t)=\omega(t)dt=\omega d\xi(t)$ also denotes Wiener noise increment satisfying $\overline{d\xi(t)}=0$ and $\overline{d\xi(t)^2}=dt$.\par
To solve the SDEs, it is necessary to properly describe the initial state. In the TWM, this is achieved by random sampling according to the probability distribution given by the Wigner function of the initial state. Specifically, for polar state with $N$ atoms, the Wigner function is found to be
\begin{equation}
W(\psi_1, \psi_1^*, \psi_0, \psi_0^*, \psi_{-1}, \psi_{-1}^*)|_{t=0}=\left(\dfrac{2}{\pi}\right)^3\exp(-2|\psi_1|^2)\exp(-2|\psi_{-1}|^2)\exp(-2|\psi_0-\sqrt{N}|^2).
\end{equation}
In numerical simulation, we typically choose $10^5$ trajectories with initial conditions
\begin{equation}
\psi_1=\dfrac{1}{2}(\alpha_1+i\beta_1), \quad\psi_{-1}=\dfrac{1}{2}(\alpha_{-1}+i\beta_{-1}), \quad\psi_0=\sqrt{N}+\dfrac{1}{2}(\alpha_0+i\beta_0),
\label{Wigner}
\end{equation}
where $\alpha_i$ and $\beta_i$ are real numbers independently sampled from standard normal distribution. The average and variance of mode population are then obtained from the average of these trajectories, according to $\langle{N}_i\rangle=\overline{\psi_i^*\psi_i}-1/2$, $(\Delta {N}_i)^2=(\Delta\psi_i^*\psi_i)^2-1/4$.\par

%
%
%
%

\clearpage
\end{widetext}

\begin{figure}[t]
	\center
	\includegraphics[width=1\linewidth]{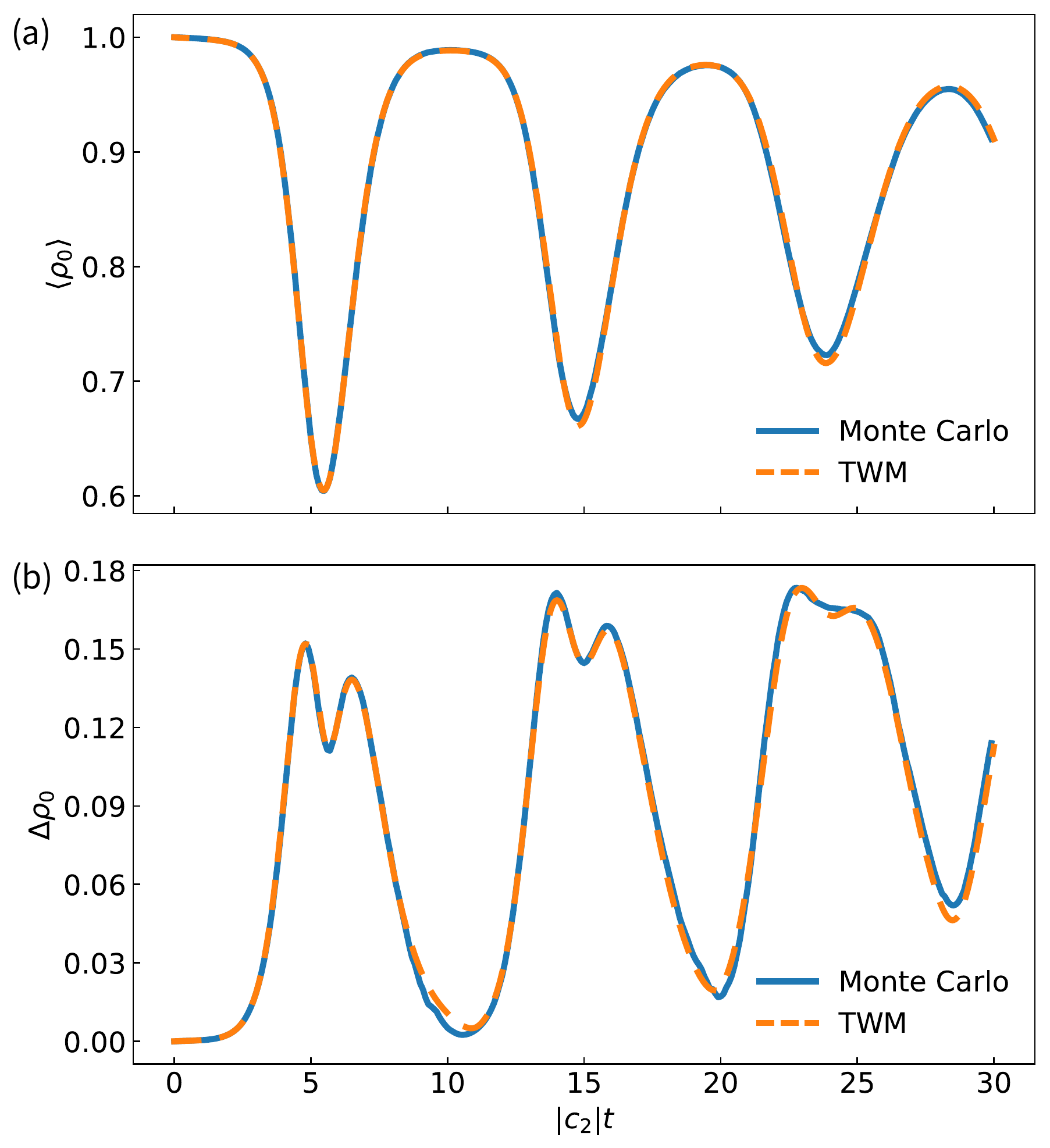}
	\caption{The comparison for the simulations based on solving master equation with Monte-Carlo (MC) wave function approach (blue line) and SDEs (orange dashed line) verifies the accuracy of the TWM. Calculations are performed for a system with 10000 atoms at $q=|c_2|$, loss rate of $0.001|c_2|$ and RF noise strength $\omega=0$. The mean value and standard deviation of $\rho_0$ are obtained from the statistics of 10000 (1000) trajectories for TWM (Monte-Carlo) simulations.}
	\label{twa_vs_mc}
\end{figure}

To confirm the validity of TWM, we compare the dissipative spin mixing dynamics starting with all atoms in $|1,0\rangle$ spin component by using TWM and Monte-Carlo (MC) wave function simulations. As shown in Fig.~\ref{twa_vs_mc}, the mean value and standard deviation of $\rho_0$ from these two approaches agree well for extended time evolutions. Given the high computation efficiency of the TWM approach for large systems with respect to MC, we mainly employ it for the theoretical simulations reported in the main text.

\section{Phase Sensitivity}\label{sens}
\subsection{Standard Quantum Limit}\label{sec_sql}
To discuss the metrological gain of a nonlinear interferometer, we need to properly specify its classical reference, namely the standard quantum limit (SQL). In the traditional two-mode Ramsey interferometry, SQL is known to be $1/\sqrt{N}$ for $N$ uncorrelated input particles. However this value depends on the types of interferometers employed. For example, the SQL of a $M$-mode Ramsey interferometer is $1/(M-1)\sqrt{N}$~\cite{Zou:2018aa}. A general definition for SQL is dependent on the optimal phase sensitivity that can be achieved with uncorrelated particles under the \textit{same} phase sensing protocol. It is straight-forward to check that the above two cases indeed satisfy this definition. Thus the specific SQL for the nonlinear interferometer discussed here follows the same spirit. With the phase imprinting operation $ U(\phi)=\exp(i\phi N_0/2)$, the phase sensitivity achieved by a classical pure probe state $|\psi\rangle$ is limited by the quantum Cramer-Rao bound (QCRB)~\cite{Braunstein:1994aa} $\Delta\phi_{\rm{QCRB}}=1/\sqrt{F_Q[ N_0/2, |\psi\rangle]}=1/(\Delta{ N}_0)=1/\sqrt{N\langle{\rho}_0\rangle(1-\langle{\rho}_0\rangle)}$, where $F_Q=(\Delta{ N}_0)^2$ denotes the quantum Fisher information (QFI), and ${\rho}_0={N}_0/N$ refers to the fractional population in the $|1, 0\rangle$ mode. This bound reaches its minimum of
\begin{equation}
\Delta\phi_{\rm{SQL}}=2/\sqrt{N},
\label{sql}
\end{equation}
at $\langle{\rho}_0\rangle=0.5$, and is therefore defined as the SQL of our nonlinear interferometer. We note that for $1-\langle{\rho}_0\rangle\ll 1$, the QCRB approximately reduces to $1/\sqrt{N(1-\langle{\rho}_0\rangle)}=1/\sqrt{\langle{ N}_1+{ N}_{-1}}\rangle$, which recovers the SQL of the atomic SU(1,1) interferometer in the undepleted pump regime~\cite{Linnemann:2016aa}. 

\subsection{Phase estimation}
In our experiment, the spinor phase is extracted by measuring the population in $|1,0\rangle$ component.
The corresponding phase sensitivity is lower bounded by the Cramer-Rao bound (CRB) $\Delta\phi_{\rm{CRB}}=1/\sqrt{F_C(\phi)}$,
 where $F_C(\phi)$ is the classical Fisher information (CFI) defined as~\cite{Gabbrielli:2015aa}
\begin{equation}\label{FC}
F_C(\phi)=\sum_{\rho_0}\frac{1}{P(\rho_0|\phi)}\bigg[\frac{dP(\rho_0|\phi)}{d\phi}\bigg]^2,
\end{equation}
with $P(\rho_0|\phi)$ being the probability distribution function of ${\rho}_0$.
In the TWM approach, numerical calculation of CFI using this formula tends to diverge when $P(\rho_0|\phi)$ approaches $0$. So we adopt the method based on the Hellinger distance~\cite{Strobel:2014aa} instead, which is defined as
\begin{equation}
d_H^2(\phi, \phi_0)=\dfrac{1}{2}\sum_{\rho_0}\left[\sqrt{P(\rho_0|\phi)}-\sqrt{P(\rho_0|\phi_0)}\right]^2.
\end{equation}
It is related to $F_C(\phi)$ in the vicinity of $\phi=\phi_0$ according to
\begin{equation}
d_H^2(\phi, \phi_0)=\frac{F_C(\phi)}{8}(\phi-\phi_0)^2 + O((\phi-\phi_0)^3).
\label{cfi}
\end{equation}
Therefore $F_C(\phi)$ can be extracted by a parabolic fitting of $d_H^2(\phi, \phi_0)$.

\begin{figure}[t]
	\centering
	\includegraphics[width=1\linewidth]{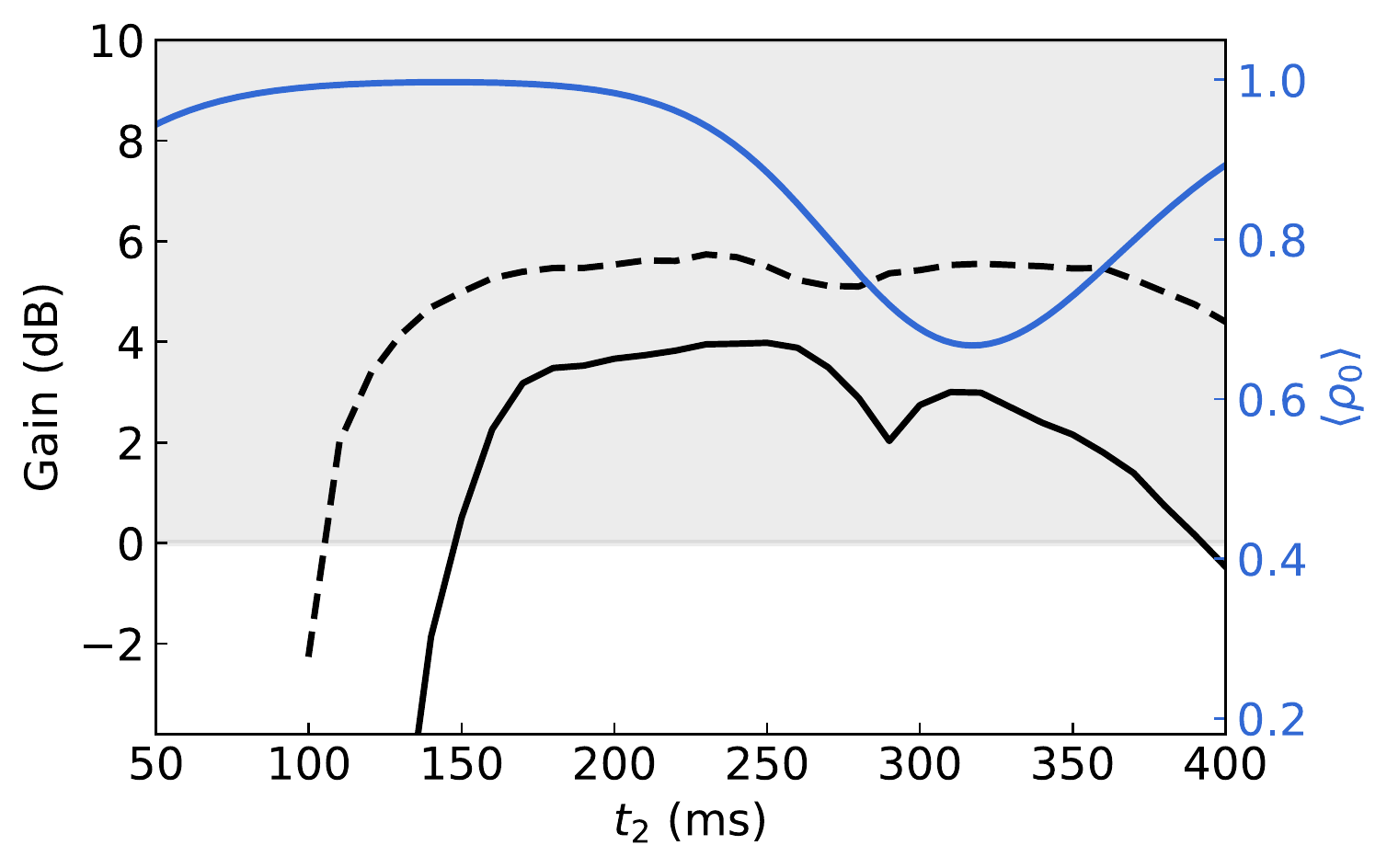}
	\caption{Comparison of metrological gains obtained by measuring $\langle\rho_0\rangle$ and $\Delta\rho_0$ (black solid line) with the CRB (black dashed line). Simulations are carried out by solving Eq.~\eqref{rfsde} with $10^5$ trajectories. CFI is extracted by fitting the Heillinger distance with parabolic function over a phase range of $[-2.8,2.8]~\rm{mrad}$. Other parameters are typical experimental settings for atom number $N=26500$, $c_2=-2\pi\times3.85$ Hz, $q=2\pi\times3.8$ Hz, loss rate $\gamma=0.09~\rm{s^{-1}}$ and RF noise strength $\omega=2\pi\times0.017~\sqrt{\rm{mHz}}$.}
	\label{sup_Fig3}
\end{figure}
In Fig.~\ref{sup_Fig3}, we compare the metrological gains obtained by measuring the mean value and standard deviation of $\rho_0$ (black solid line) with the CRB (black dashed line). The spin mixing time for `path' splitting $t_1$ is fixed at $290$~ms and the gains (or CRB) are optimized over $\phi\in[0,0.04]$ rad. We find that for the experimental case with `path' recombining time $t_2=230$ ms, the achieved gain is two decibels short of the CRB.
We also note at the first maximum of $\langle\rho_0\rangle$, measuring the first two moments of $\rho_0$ is insufficient to characterize the variation of state with imprinted phase, although the corresponding CRB is beyond the SQL.

\bibliography{ref_SM}